\documentclass[a4paper,12pt]{article}

\usepackage{epsfig}
\usepackage{a4}
\usepackage{latexsym, amssymb}

\newcommand{\lapprox}{%
\mathrel{%
\setbox0=\hbox{$<$}
\raise0.6ex\copy0\kern-\wd0
\lower0.65ex\hbox{$\sim$}
}}
\newcommand{\gapprox}{%
\mathrel{%
\setbox0=\hbox{$>$}
\raise0.6ex\copy0\kern-\wd0
\lower0.65ex\hbox{$\sim$}
}}

\newcommand{\ba}{\begin{array}}
\newcommand{\ea}{\end{array}}
\newcommand{\bd}{\begin{displaymath}}
\newcommand{\ed}{\end{displaymath}}
\newcommand{\be}{\begin{equation}}
\newcommand{\ee}{\end{equation}}
\newcommand{\bea}{\begin{eqnarray}}
\newcommand{\eea}{\end{eqnarray}}

\def\fb{\, {\rm fb}}

\def\met{E_T \hspace*{-1.1em}/\hspace*{0.5em}}
\def\lv{L \hspace*{-0.4em}/\hspace*{0.3em}}
\def\tev{\, \, {\rm TeV}}
\def\gev{\, \, {\rm GeV}}

\def\neu {{\tilde {\chi_1}}^0}
\def\cha {{\tilde {\chi_1}}^{\pm}}

\def\lslep {\tilde{e_L}}
\def\stau {\tilde{\tau_1}}
\def\stop {\tilde{t_1}}
\def\sbo {\tilde{b_1}}

\def\mel {m_{\lslep}}
\def\mneu {m_{\neu}}
\def\mcha {m_{\cha}}
\def\mstau {m_{\stau}}

\def\thefootnote{\fnsymbol{footnote}}

\begin{document}

\begin{titlepage}

\begin{flushright}
{\small 
November 02, 2011 \\ 
RECAPP-HRI-2011-006}
\end{flushright}

\vspace*{0.2cm}
\begin{center}
{\Large {\bf Same-sign trileptons at the LHC: a window to
    lepton-number violating supersymmetry}}\\[2cm]

Satyanarayan Mukhopadhyay\footnote{satya@hri.res.in} and 
Biswarup Mukhopadhyaya\footnote{biswarup@hri.res.in}, 
 \\[0.5cm]  

{\it Regional Centre for Accelerator-based Particle Physics \\
Harish-Chandra Research Institute \\
Chhatnag Road, Jhusi \\ 
Allahabad - 211 019, India}\\[2cm]
\end{center}

\begin{abstract}
We present a detailed investigation to establish that lepton-number
(L) violating supersymmetry (SUSY) can be effectively probed at the
LHC in the practically background-free same-sign trilepton (SS$3
\ell$) and same-sign four-lepton (SS$4 \ell$) channels. With this in
view, we extend our earlier analysis of SS$3 \ell$ and SS$4 \ell$
signals by considering situations based on minimal supergravity as
well as a phenomenological SUSY model.  We find that the R-parity
violating scenario predicts large event rates, for both the 7 and 14
TeV runs. Furthermore, we show that it is extremely unlikely to ever
achieve similar rates in R-parity conserving SUSY. In addition, we
show how SS$3 \ell$ and SS$4 \ell$, in conjunction with the mixed-sign
trilepton and four-lepton channels, can be used to extract dynamical
information about the underlying SUSY theory, namely, the Majorana
character of the decaying lightest neutralino and the nature of
L-violating couplings. We define suitable variables and relationships
between them which can be verified experimentally and which are
largely independent of the SUSY production cross-sections and the
cascade decay branching fractions.  These theoretical predictions are
validated by Monte Carlo simulations including detector and background
effects.
\end{abstract}

\pagestyle{plain}

\end{titlepage}


\setcounter{page}{1}

\renewcommand{\thefootnote}{\arabic{footnote}}
\setcounter{footnote}{0}

\section{Introduction}
\label{sec:intro}

Discovering new physics at the Large Hadron Collider (LHC) is a
cherished dream. In this context, it is always useful to isolate
signals which are distinctive of specific new scenarios on the one
hand, and are less background-prone on the other.  Final states
containing a multitude of leptons undoubtedly satisfy the second
criterion. In order to address the first criterion through them, it is
often necessary to probe additional features of the leptons. One such
feature is the sign(s) of the leptonic charge(s). It is
well-established now that same-sign dilepton (SSD) carries a rather
distinct signature of supersymmetry (SUSY)~\cite{SSD}, and other new
physics scenarios~\cite{Goran}, once we carefully apply the event
selection criteria to suppress the top-antitop background.

A curiosity that immediately arises is whether same-sign leptons of
higher multiplicity can tell us something more.  Although this idea of
same-sign trileptons (SS$3 \ell$) was floated originally in the
context of top quark signals \cite{Barger-top}, its efficacy in new
physics search was unexplored until very recently. This is somewhat
unfortunate, because the standard model (SM) backgrounds for them are
extremely small.  Some studies in the context of heavy neutrino
signals were reported, though with rather limited
scope~\cite{Ozcan}. In a more recent work, we pointed out that SS$3
\ell$ as well as its four-lepton extension (SS$4 \ell$) had
considerable potential in unearthing scenarios where $Z_2$-type
discrete symmetries were broken in a limited manner~\cite{SMBM-1}. In
particular, we showed that various R-parity violating SUSY
scenarios~\cite{Barbier} (with R = $(-1)^{(3B + L + 2S)}$, $B, L$ and
$S$ being baryon number, lepton number and spin, respectively)
predicted large signal rates for SS$3 \ell$ and moderate rates for
even SS$4 \ell$, with hardly any backgrounds. The SS$3 \ell$ signal is
substantial over a range of the parameter space in the 7 TeV run,
while the predictions for both SS$3 \ell$ and SS$4 \ell$ are copious
for 14 TeV. This suggestion has since been utilized in a number of
subsequent studies~\cite{SS3l-Others}. In this paper, we present a
more extensive study in the same direction, pointing out a number of
new possibilities of the SS$3 \ell$ signal.

Let us begin with some explanation of why such a study is relevant.
First of all, the LHC searches for new physics, particularly SUSY, at
the initial stage, are concentrating on signals with large missing
transverse energy. So far the results have been negative.  If they
continue to be so, some possibilities to consider will be (a) SUSY
without R-parity, (b) a highly compressed SUSY spectrum, and (c) SUSY
with stable visible particles. While the signatures of each of the
above scenarios have been proposed and investigated in the literature,
the SS$3 \ell$ and SS$4 \ell$ signals are exclusively indicative of
SUSY without R-parity, with lepton number violation. Since such
signals can arise with large rates even during the early run, they are
worth studying seriously, from the sheer event counting point of view.

L-violating SUSY has considerable appeal, because mechanisms of
neutrino mass generation are suggested there.  It is also being
increasingly realised nowadays that one may end up with a dark matter
candidate such as the axino or the gravitino in spite of R-parity
violation. Some search limits for R-parity violating SUSY exist in the
literature, based on multilepton ($\ge 3 \ell$) signals. However, SS$3
\ell$ is a rather more unequivocal indication of R-parity violation,
since it is very difficult to produce three leptons of the same sign
unless the seed of lepton number violation is there.  Moreover, as
will be discussed later in this paper, enhanced rates for such signals
are very unlikely to be found in R-parity conserving versions of SUSY,
even in a purely phenomenological scan of its parameter space. The
background is also vanishingly small, in contrast with the other
channels advocated so far. With this in view, we also demonstrate
regions in the parameter space where one can have five signal events
with zero (in practice, $\le 1$) background events for some
luminosity.

The enhanced signal rates were predicted in our earlier study within
the framework of a minimal supergravity (mSUGRA) scenario.  It is,
however, important to go beyond the most simple of `top-down' models
and investigate SUSY signals at the LHC in a phenomenological,
`bottom-up' approach. In the current work, we have taken such an
approach and looked at the SUSY parameter space in a relatively
unbiased manner, although some simplification has been inevitable in
order to keep the number of free parameters manageable. This allows us
to point out features of the SUSY spectrum, for which same-sign
multileptons are most likely to be observed.

We have opened another new direction in the present study. There are a
number of ways in which R-parity can be violated via L, since one can
have the so-called $\lambda$-type, $\lambda^{'}$-type and the
L-violating bilinear terms in the superpotential. Besides, the
lightest neutralino need not be the lightest SUSY particle (LSP) when
R-parity is violated, the stau-LSP scenario being the most common
possibility.  We contend here that the SS$3 \ell$ and SS$4 \ell$
signal rates, in conjunction with their mixed-sign counterparts of the
same multiplicity, display certain mutual relations which distinguish
among at least some of the candidate scenarios.  And these relations
are largely independent of the detailed information of the SUSY
cascades. Consequently, one may use these signals to find out in a
generic way the distinction among the $\lambda$-type,
$\lambda^{'}$-type or bilinear couplings.  Although our discussion is
largely based on scenarios with a neutralino LSP, alternative
scenarios, for example, with stau LSP, can be brought within its
scope, as has been briefly indicated in the paper.

It should be noted that same-sign multileptons in general, and SS$3
\ell$ in particular, can be seen in some other non-standard scenarios
as well. In most cases, however, the rates are considerably smaller
than what one would expect for R-parity violating cases with new
particles with similar masses.  The first example of this is minimal
SUSY standard model (MSSM) where R-parity is conserved; our scan of
its parameter space, with the usual constraints satisfied, reveals
rather low event rates for SS$3 \ell$.  One has predictions of some
interest for Little Higgs theories where T-parity is broken by the
Wess-Zumino-Witten anomaly terms~\cite{Littlest_Higgs,Hill_Hill,LHT}.
However, it has been shown in our previous study that the rates are
much smaller than those for R-parity violating SUSY with a spectrum of
similar masses~\cite{SMBM-1}. In addition, models with heavy charged
leptons and Majorana neutrinos~\cite{Ozcan} and triply charged heavy
leptons~\cite{Picek} can also lead to an SS$3 \ell$ signature.

The paper is organized as follows. Section 2 is devoted to the
standard model contributions to same-sign and mixed-sign multilepton
channels, and we suggest event selection criteria that suppress such
contributions as potential backgrounds to the new physics signals.  In
section 3, we review the different cases of R-parity violating SUSY,
and show the event rates for different benchmark points for mSUGRA,
for both the 14 and 7 TeV runs. Section 4 contains a study where the
parameters are varied in a more phenomenological manner, and regions
where R-parity violating SUSY shows up in the SS$3 \ell$ channel are
pointed out. We also explain in the same section why the SS$3 \ell$
and SS$4 \ell$ signals are not expected to occur with appreciable
rates when R-parity is conserved, even in a generic MSSM model. In
section 5, we show how we can extract information on the Majorana
character of the lightest neutralino and the dynamics of R-parity
violation from SS$3 \ell$ and SS$4 \ell$ signals. We summarise and
conclude in section 6.

\section{Standard model backgrounds}
As we have discussed in Ref.~\cite{SMBM-1}, the SM backgrounds to SS$3
\ell$ are vanishingly small.  Though the channels $t\bar{t}$, $t
\bar{t} W$, $t \bar{t} b \bar{b}$ and $t \bar{t} t \bar{t}$ can give
rise to such events, the only appreciable contribution after various
kinematic cuts comes from $t \bar{t} W$. These cuts are designed
mainly to suppress the leptons coming from semi-leptonic bottom and
charm decays~\cite{Sullivan}.

We select events with three and only three leptons in the signal (for
SS$3 \ell$), all of which have to be of the same-sign. In addition, we
demand the following basic selection criteria:

\begin{enumerate}
 \item $p_{T}^{l_{1}}> 30 \gev$, $p_{T}^{l_{2}}> 30 \gev$, $p_{T}^{l_{3}}>
  20 \gev$, where $l_1$, $l_2$ and $l_3$
  are the three leptons ordered according to their $p_T$'s

\item Missing transverse energy, $\met > 30 \gev$ (in order to reduce
  events with jets faking as leptons).

\item Lepton rapidity $|\eta| < 2.5$.

\item Lepton-lepton separation $\Delta R_{ll} \ge 0.2$, where $(\Delta
  R)^2 = (\Delta \eta)^2 + (\Delta \phi)^2$ quantifies the separation
  in the pseudorapidity-azimuthal angle plane.

\item Lepton-jet separation $\Delta R_{lj} \ge 0.4$ for all jets with $E_T \ge 20$ GeV.

\item Relative isolation criterion to restrict the hadronic activity
  around a lepton has been used, i.e., we demand $\sum p_T$ (hadron)
  /$p_T$ (lepton)$\le 0.2$, where the sum is over all hadrons within a
  cone of $\Delta R \le 0.2$ around the lepton.

\item Electron and muon selection efficiencies were taken to be $70
  \%$ and $90 \%$ respectively~\cite{TDR}.

\end{enumerate}

The events for the above-mentioned SM background processes
contributing to SS$3 \ell$ were generated with the code
ALPGEN~\cite{ALPGEN}, and showering, decays and hadronisation were
done using PYTHIA 6.421~\cite{PYTHIA}. The effect of $B^0 - \bar{B}^0$
mixing on lepton signs has been taken into account within PYTHIA. We
have approximated the detector resolution effects by smearing the
energies (transverse momenta) of the leptons and jets with Gaussian
functions~\cite{TDR,GMM}.  After imposing the above cuts the total
SS$3 \ell$ contribution from the SM at $14 \tev$ LHC turns out to be
$2.50 \times 10^{-3} \fb$, of which $2.44 \times 10^{-3} \fb$ comes
from the $t \bar{t} W$ process. At $7 \tev$ LHC, the total SM
cross-section for SS$3 \ell$ comes down to $7.01 \times 10^{-4}
\fb$. We have used CTEQ6L1~\cite{CTEQ} parton distribution functions
for all our signal and background calculations. As mentioned before,
for further details on these backgrounds we refer the reader to our
previous study in Ref.~\cite{SMBM-1}.

Needless to say, the SM backgrounds to the SS$4 \ell$ channel will be
even smaller than SS$3 \ell$, and can be safely neglected. As we shall
see later in section~\ref{xy-var}, we can construct certain
observables which depend only on the Majorana nature of the LSP and
the L-violating coupling involved and not upon other parameters
determining the cascade decays. In addition to the SS$3 \ell$ and SS$4
\ell$ cross-sections, these variables shall also depend upon the total
trilepton ($3 \ell$) and four-lepton ($4 \ell$) cross-sections in a
given scenario, with specific kinematic criteria. Thus we need to
evaluate and include the SM backgrounds in the $3 \ell$ and $4 \ell$
channels, and subject them to the same set of cuts irrespective of the
sign of leptons.  Therefore, while a Z-veto (removing events
containing same flavour, opposite-sign leptons with invariant mass
around the Z-boson mass) is often used to reduce the SM backgrounds in
the sign-inclusive $3 \ell$ and $4 \ell$ channels, we cannot use such
a veto here. Also, we use kinematic variables that only depend upon
the lepton $p_T$'s and the $\met$ in an event. We find it useful to
select events in terms of the variables $m_{eff}^{\ell}$ and
$m_{eff}$, defined as follows:

\begin{eqnarray} \label{M_eff}
m_{eff}^{\ell}=\sum p_{T}^{leptons}\\
m_{eff}=\sum p_{T}^{leptons} + \met,  
\end{eqnarray}
where the missing transverse energy is given by 
\be \label{missing_E_T}
\met =\sqrt{\left(\sum p_x \right)^2+\left(\sum p_y\right)^2}. 
\ee
Here the sum goes over all the isolated leptons, the jets, as well as
the `unclustered' energy deposits.
\begin{table}[htb]
\begin{center}
\begin{tabular}{|l|c|c|c|c|c|}
\hline
Cut & $W^\pm (Z_0 / \gamma^{\star})$ & $t \bar{t}$ & $t \bar{t} (Z_0 / \gamma^{\star})  $ &$t \bar{t} W^\pm$ & Total \\
\hline
Basic cuts &34.50  &9.88 &2.82 & 0.73 &47.93\\
\hline
$m_{eff}^{\ell}>100\,\gev$&33.53  &8.23 & 2.80&0.71 &45.27 \\
$m_{eff}^{\ell}>200\,\gev$&5.01   &0.00 & 1.43&0.33 &6.77 \\
\hline
$m_{eff}>150\,\gev$       &32.06  &8.23 & 2.80&0.72 &43.81 \\
$m_{eff}>250\,\gev$       &6.18   &1.65 & 1.81&0.48 &10.12 \\
\hline 
\end{tabular}
\end{center}
\caption{\label{3L} SM contributions to the trilepton channel at $14
  \tev$ LHC.  The $m_{eff}^{\ell}$ or $m_{eff}$ cut is applied one at
  a time.  All the cross-sections are in femtobarns. }
\end{table}

\begin{table}[htb]
\begin{center}
\begin{tabular}{|l|c|c|c|}
\hline
Cut & $(Z_0 / \gamma^{\star}  )(Z_0 / \gamma^{\star}  )$ & $t \bar{t} (Z_0 / \gamma^{\star})$ & Total \\
\hline
Basic cuts &9.33   &0.46 &9.79\\
\hline
$m_{eff}^{\ell}>100\,\gev$&9.25   &0.46 &9.71 \\
$m_{eff}^{\ell}>200\,\gev$&3.71    &0.32 &4.03 \\
\hline
$m_{eff}>150\,\gev$       &7.87   &0.45 &8.32 \\
$m_{eff}>250\,\gev$       &1.67   &0.36 &2.03 \\
\hline 
\end{tabular}
\end{center}
\caption{\label{4L} SM contributions to the four-lepton channel at $14
  \tev$ LHC.  The $m_{eff}^{\ell}$ or $m_{eff}$ cut is applied one at
  a time. The $t \bar{t}$ contribution is zero after the basic lepton
  selection and isolation cuts.  All the cross-sections are in
  femtobarns. }
\end{table}

The SM contributions coming from the sign-inclusive $3 \ell$ and the
$4 \ell$ channels are shown in Tables~\ref{3L} and~\ref{4L}
respectively. The predictions for the 14 TeV run only are presented
here; although the predictions for 7 TeV, too, are very small, we do
not expect enough statistics for performing our suggested analysis
there.  We show the cross-sections after different cuts on the
$m_{eff}^{\ell}$ and $m_{eff}$ variables. In the $3 \ell$ channel the
major backgrounds are $W^\pm (Z_0 / \gamma^{\star})$, $t \bar{t}$, $t
\bar{t} (Z_0 / \gamma^{\star}) $ and $t \bar{t} W^\pm$. Here, the
$W^\pm (Z_0 / \gamma^{\star})$ and $t \bar{t} (Z_0 / \gamma^{\star}) $
processes include the effect of $Z$, $\gamma^*$ and their
interference. In the $4 \ell$ channel, the dominant SM contributions
come from $(Z_0 / \gamma^{\star} )(Z_0 / \gamma^{\star} )$, $t
\bar{t}$ and $t \bar{t} (Z_0 / \gamma^{\star})$. The processes $W^\pm
(Z_0 / \gamma^{\star})$ and $t \bar{t} (Z_0 / \gamma^{\star})$ were
simulated using MadGraph 5~\cite{MG5} and PYTHIA, $t \bar{t} W^\pm$
using ALPGEN and PYTHIA and $t \bar{t}$ and $(Z_0 / \gamma^{\star}
)(Z_0 / \gamma^{\star} )$ using PYTHIA alone. For the $t \bar{t}$
process we have multiplied the leading order cross-section from PYTHIA
by a K-factor of 2.2 according to the analysis in
Ref.~\cite{Cacciari_et_al}. We have taken showering, hadronisation and
multiple interaction effects into account in all of our simulations.

\section{SS$3 \ell$ in L-violating SUSY: a brief review of the different cases}
In this section, we shall very briefly describe how SS$3 \ell$ can
arise in different scenarios of lepton-number (L) violating
supersymmetry. For a detailed discussion on this, we refer the reader
to our previous work on this subject~\cite{SMBM-1}. \\

The superpotential in R-parity violating SUSY can contain the
following $\Delta L = 1$ terms, over and above those present in the
MSSM:
 
\begin{equation}
\nonumber
 W_{\lv} = \lambda_{ijk}L_iL_j{\bar{E}}_k 
+ {\lambda^{\prime}_{ijk}}L_iQ_j{\bar D}_k+ {\epsilon}_i L_i H_2
\end{equation}

{\bf {\em Case 1}:} With the $\lambda$-type terms, we consider two
possibilities, namely, having (a) the lightest neutralino ($\neu$) and
(b) the lighter stau ($\stau$) as the lightest SUSY particle (LSP). In
(a), SS$3 \ell$ can arise if $\neu$ decays into a neutrino, a tau
($\tau$) and a lepton of either of the first two families.  When the
$\tau$ decays hadronically, the two leptons from two $\neu$'s produced
at the end of SUSY cascades are of identical sign in 50\% cases. An
additional lepton of the same sign, produced in the decays of a
chargino ($\cha$) in the cascade, leads to SS$3 \ell$. If there is
just one $\lambda$-type coupling (we have used $\lambda_{123}$ for
illustration), there is no further branching fraction suppression in
LSP decay, and one only pays the price of $\cha$-decay into a lepton
of the same sign. In (b), two same-sign $\stau$'s can be produced from
two $\neu$'s, due to its Majorana character. Each of these $\stau$'s
goes into a lepton and a neutrino; these two leptons, together with
one of identical sign from the cascade, lead to SS$3 \ell$ signals.

{\bf {\em Case 2}:} With $\lambda'$-type interactions, a $\neu$-LSP
decays into two quarks and one charged lepton or neutrino. If the LSP
is not much heavier than the top quark, and if the effect of the
difference between up and down couplings of the neutralino can be
neglected, we obtain SSD's from a pair of $\neu$'s roughly in 12.5\%
of the cases. If another lepton of the same sign arises from a $\cha$,
SS$3 \ell$ is an immediate consequence. Therefore, the overall rate of
SS$3 \ell$ can be substantial in this case as well. Here, (and also
partially in case 1(b)), the large boost of the $\neu$ can lead to
collimated jets and leptons, making the latter susceptible to
isolation cuts.

Although most of the analysis we have presented for such couplings is
based on a $\neu$-LSP scenario, we shall see later that one with
$\stau$-LSP, too, has potential for SS$3 \ell$ events.

{\bf {\em Case 3}:} With bilinear R-parity breaking terms ($\sim
\epsilon_i$), the most spectacular consequence is the mixing between
neutralinos and neutrinos as well as between charginos and charged
leptons.  Consequently, over a substantial region of the parameter
space, a $\neu$ LSP in this scenario decays into $W \mu$ or $W \tau$
in 80\% cases altogether, so long as the R-parity breaking parameters
are in conformity with maximal mixing in the $\nu_{\mu} - \nu_{\tau}$
sector~\cite{R_M_2}. From the decay of the two $\neu$'s, one can
obtain SSD's either from these $\mu$'s, or from the leptonic decay of
the $W$'s or the $\tau$'s. An additional lepton from the SUSY cascade
results in SS$3 \ell$ again. Adding up all the above possibilities,
the rates can become substantial.

Again, in addition to a $\neu$-LSP, a $\stau$-LSP also can lead to the
signals under consideration. We shall briefly mention such
possibilities later in our discussion.

\subsection{mSUGRA benchmark points and rates for  $14 \tev$}

\begin{figure}[htb!]
\begin{center}
\centerline{\epsfig{file=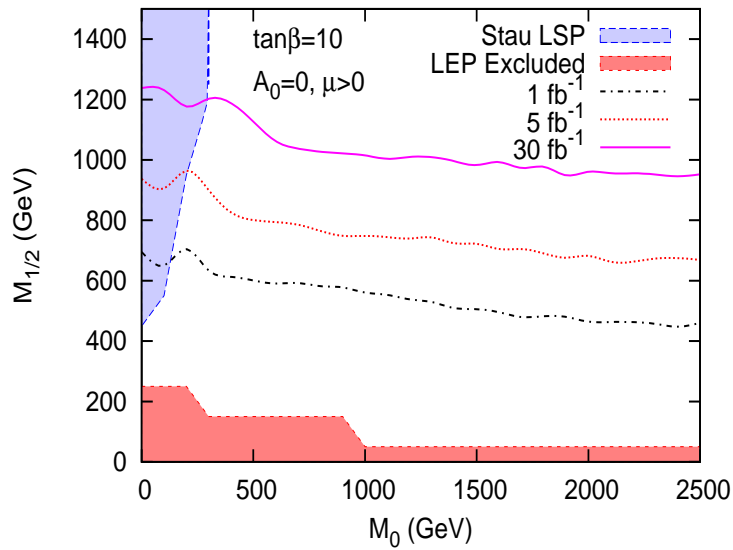,width=10cm,height=8.0cm,angle=-0}} 

\caption{(Color online) 5-events LHC reach with SS$3 \ell$ in the
  $M_0-M_{1/2}$ plane for R-parity violating mSUGRA, at $\sqrt{s}=14
  \tev$, with $\lambda_{123}=0.001$, after all selection cuts.}
\label{scan}
\end{center}
\end{figure}

In ~\cite{SMBM-1}, the SS$3 \ell$ cross-sections for a few
representative points from the mSUGRA parameter space were presented,
for both the $7 \tev$ and $14 \tev$ runs.  Those benchmark points,
corresponding to the various cases discussed in the previous
sub-section, are shown in Table~\ref{tab2}. We show the values of
$M_0$ and $M_{1/2}$ ($M_0$ and $M_{1/2}$ being respectively the
universal scalar and gaugino mass at high scale), the ratio of the
vacuum expectation values of the two Higgs doublets $\tan \beta$, as
well as the values of other relevant SUSY parameters at the
electroweak scale (fixed here at $\sqrt{m_{{\tilde t}_1}m_{{\tilde
      t}_2}}$, where ${\tilde t}_1$ and ${\tilde t}_2$ are the two
mass eigenstates of the top squarks). We use fixed values for the
other mSUGRA parameters, namely, the universal soft-breaking trilinear
scalar interaction $A_0=0$ and the Higgsino mass parameter $\mu>0$.
Since the values of the L-violating couplings are very small, they do
not affect the renormalisation group running of mass parameters from
high to low scale~\cite{Allanach}. We have therefore generated the
spectrum using SuSpect 2.41~\cite{SUSPECT} and interfaced it with
SDECAY~\cite{SDECAY} by using the programme SUSY-HIT~\cite{SUSY-HIT}
(for calculating the decay branching fractions of the sparticles) and
finally have interfaced the spectrum and the decay branching fractions
to PYTHIA, which is used to generate all possible SUSY production
processes. Also, we have neglected the role of R-violating
interactions in all stages of cascades excepting when the LSP is
decaying. The value of each trilinear coupling ($\lambda, \lambda'$)
used for illustration is 0.001. For case 3, The values of the
$\epsilon$-parameters are chosen consistently with the neutrino data;
essentially, they are tuned to sneutrino vacuum expectation values of
the order of 100 keV, in a basis where the bilinear terms are rotated
away from the superpotential. The values of $\epsilon_i$ are also of
this order in the absence of any additional symmetry. The exact values
of $\epsilon_i$ that correspond to points $3(1)$ and $3(2)$ in
Table~\ref{tab2} depend also on other parameters of the model, such as
the L-violating soft terms in the scalar
potential~\cite{Datta_Mukho_Roy}. However, the range of values of
these parameters is of little consequence to the neutralino decay
branching ratios. Therefore, with appropriate values of these soft
terms, $\epsilon_3\approx 100$~keV, $\epsilon_1=\epsilon_2=0$ is
consistent with all our results. In order to demonstrate the reach of
the LHC in the SS$3 \ell$ channel, we show in Figure~\ref{scan}, the
boundary contours of regions in the $M_0-M_{1/2}$ plane, where at
least 5 signal events (with zero background event expected) can be
obtained with a given integrated luminosity. This scan was performed
for a sample case (case 1) with fixed values for the other mSUGRA
parameters ($\tan \beta=10, A_0=0,\mu>0$). Similar discovery reaches
are expected for the other cases also. This plot, which we also
presented in Ref.~\cite{SMBM-1}, has now been improved by using proper
interpolation routines for points lying in between the simulated grid
in the $M_0$-$M_{1/2}$ plane. In addition, in Figure~\ref{scan} we
have used a 5 signal events discovery criterion while in
Ref.~\cite{SMBM-1} we used a 10 signal events discovery
criterion. Since we do not expect any background event in the SS$3
\ell$ channel even with $30 \fb^{-1}$ integrated luminosity, 5 signal
events should be sufficient for a discovery. Note that there is a
sharp fall observed in each curve of Figure~\ref{scan}. As we increase
$M_0$ for a given $M_{1/2}$, the first two family sleptons eventually
become heavier than the chargino, thereby reducing the branching
fraction of $\cha\rightarrow l^{\pm} \nu \neu$. This leads to a drop
in the SS$3 \ell$ cross-section, giving rise to the faster fall in the
curves.

We refer the reader to \cite{SMBM-1} for the total signal
cross-sections for the various other cases of R-parity violation. It
is clear that, with at least $30 \fb^{-1}$ of integrated luminosity,
they can enable one to perform the analysis suggested in later
sections for extracting the exact dynamics of R-parity violation.

\begin{table}[htb!]
	\centering

\begin{tabular}{|c |c|c|c| c| c| c| c| c| c|}

\hline
Case &$M_0$&$M_{1/2}$&$\tan \beta$ & $m_{\tilde {g}}$ &$\mcha$&$\mneu$&$\mstau$& $\mel$&RPV\\ 
   &(GeV)&(GeV)& &(GeV)&(GeV)&(GeV)&(GeV)&(GeV)&Coupling \\ 
\hline
1a(1)&75 &275 & 15 &661  & 200 &$108^*$& 115 & 204& $\lambda_{123}$\\ 
1a(2)&200 &250 &40 &610 &183 & $99^*$& 139 & 265& $\lambda_{123}$\\ 
1a(3)&100 &435 &5& 1009 &331&$176^*$&191 & 309&$\lambda_{123}$ \\ 
1a(4)&300 &435 &40 & 1016 &337 &$178^*$&246&418&$\lambda_{123}$\\ 
\hline
1b(1)&0 &325 &10 & 770 & 241 & 129 & $118^*$ & 222 & $\lambda_{123}$ \\ 
1b(2)&160 &250 & 40 & 608 & 182 & 98  & $94^*$  & 236 & $\lambda_{123}$ \\ 
1b(3)&50 &435 & 5  &1008 & 330 &176  & $171^*$ & 297 & $\lambda_{123}$\\ 
1b(4)&150 &435 & 40 &1009 & 336 &178  & $109^*$ & 328 & $\lambda_{123}$ \\ 
\hline
2(1) &75 &275 & 15 &661  & 200 &$108^*$& 115 & 204&$\lambda^{\prime}_{112}$\\ 
2(2) &200 &250 &40 &610 &183 & $99^*$& 139 & 265& $\lambda^{\prime}_{112}$\\ 
2(3) &100 &435 &5& 1009 &331&$176^*$&191 & 309& $\lambda^{\prime}_{112}$ \\ 
2(4) &300 &435 &40 & 1016 &337 &$178^*$&246&418& $\lambda^{\prime}_{112}$\\ 
\hline
3(1) &100 &435 &5& 1009 &331&$176^*$&191 & 309 & $\epsilon_i$ \\ 
3(2) &300 &435 &40 & 1016 &337 &$178^*$&246&418& $\epsilon_i$ \\ 
\hline
	\end{tabular}

\caption{\label{tab2} mSUGRA benchmark points defined in
  Ref.~\cite{SMBM-1} for the various cases discussed in the text
  (e.g., 1a(1) corresponds to the first example in case 1a). The LSP
  in a given point is indicated by a * against its mass. The low-scale
  MSSM parameters were generated in an mSUGRA framework (with $A_0=0$
  and $\mu >0$). The $\lambda$ and $\lambda'$ couplings are set at
  0.001, and the $\epsilon_i$ are within the limits set by neutrino
  data. The SS$3 \ell$ cross-sections after various cuts in these
  benchmark points can be found in Ref.~\cite{SMBM-1}. }
\end{table}

\subsection{A resume of the $7 \tev$ results}
In our previous study~\cite{SMBM-1}, for the $7 \tev$ run, we
presented the points which can be discovered with an integrated
luminosity of $2 \fb^{-1}$. Since the LHC experiments are collecting
data at a fast pace and there is every chance of continuing the $7
\tev$ run upto at least $5 \fb^{-1}$, we update our $7 \tev$ results
in cMSSM to include more points which can now be accessed. In
particular, with this increase of luminosity, we find that benchmark
points with the squark-gluino masses in the $\tev$ range can also be
accessed in the SS$3 \ell$ channel. These benchmark points include
cases with $\neu$ LSP and $\lambda$-type couplings (points 1a(3) and
1a(4)) and also $\stau$ LSP with $\lambda$-type couplings (point
1b(3)). For $\neu$ LSP with $\lambda^\prime$-type couplings, the SUSY
sparticle mass-reach is somewhat smaller, and we can access masses
slightly higher than $650 \gev$ during the $7 \tev$ run (point 2(1) in
Table~\ref{tab3}), if we insist on seeing 10 signal events without
backgrounds with $5 \fb^{-1}$ of integrated luminosity. For 3 signal
events with $\le$ 1 background event, however, the reach is
considerably higher.  Thus, on the whole, the 7 TeV run has extremely
encouraging prospects for more than one R-parity violating scenarios,
from the viewpoint of total event rates.

\begin{table}[htb!]
 \centering
 \begin{tabular}{|c | c|}
\hline
Case  &$\sigma_{SS3l}$\\
                &(fb)             \\
\hline
1a(1)  &19.82 \\
1a(2)  &29.45 \\
1a(3)  &4.29 \\
1a(4)  &2.01 \\
1b(1)  &30.74 \\
1b(2)  &6.46  \\
1b(3)  &3.35  \\
2(1)   &2.07  \\
2(2)   &4.03  \\
\hline
 \end{tabular}

\caption{\label{tab3} SS$3 \ell$ cross-sections after all selection
  cuts ($\sigma_{SS3l}$) at $\sqrt{s}=7 \tev$ for the different cases
  defined in Table~\ref{tab2}. }
\end{table}

\subsection{Other possibilities in L-violating SUSY}

We now discuss some other possible cases in L-violating SUSY where one
can also get SS$3 \ell$ events. In case of a $\stau$ LSP with
$\lambda^{\prime}_{ijk}$-type couplings, the $\stau$ will directly
decay to two quarks if the index $i$ takes the value $3$. For the
other two cases where $i$ takes the value $1$ or $2$, $\stau$ cannot
decay via two-body L-violating modes. In this case, it will go through
a 4-body decay via an intermediate off-shell chargino or
neutralino. In mSUGRA type of models, the lighter stau is mostly
composed of the right-chiral field, in which case it will couple
primarily to the bino component of the neutralino. Also, the lighter
chargino there is mostly heavier than the lightest neutralino, and
therefore the propagator suppression is more for off-shell
chargino. Thus, the mode through off-shell neutralino will
dominate. Thus, we shall find the dominant decay pattern for a $\stau$
to be $\stau \rightarrow \tau {\tilde {\chi_1}}^{0(*)} \rightarrow
\tau l^\pm q q^{\prime}$. SS$3 \ell$ events arise then in a very
similar fashion as in the case of a $\neu$ LSP with
$\lambda^\prime$-type couplings.

Since the intermediate neutralino in $\stau^\pm$-decay is a Majorana
particle, we shall not have equal rates for the $\tau^{\pm} l^{+} q
q^{\prime}$ and the $\tau^{\pm} l^{-} \bar{q} \bar{q^{\prime}}$ final
states.  Also, the production of the $\stau$ in in the decay of each
neutralino has also an accompanying tau, which is another source of
leptons. Consequently, with two taus decaying together with two staus,
there is a favourable combinatoric factor for SS$3 \ell$, which
partially offsets the suppression due to branching ratios.  The exact
numerical evaluation of the relevant branching fractions and the
resulting event rates is a detailed exercise by itself.  In any case,
we expect substantial cross-sections in the SS$3 \ell$ channel, with
the usual reduction of events in the presence of $\lambda^{\prime}$
-type couplings compared to the presence of $\lambda$-type couplings.

In presence of bi-linear L-violating couplings, a $\stau$ which is the
LSP can mix with a charged Higgs, thereby leading to the decay mode
$\stau \rightarrow \tau \nu_{\tau}$, since the charged Higgs will
couple more to the tau lepton than to electrons or muons. Thus,
starting from a pair of neutralinos which can be produced in cascades,
one can obtain two same-sign tau leptons, whose further leptonic
decays can give rise to two leptons of the same sign. The third lepton
of the same sign can come in the usual way from $\cha$ decay giving
rise to SS$3 \ell$. Evidently, the rates in this case are expected to
be rather small for various branching fraction suppressions. A
detailed study of SS$3 \ell$ in the $\stau$-LSP scenario for both of
the above cases will be reported in a forthcoming publication
\cite{inprep}.

\section{SS$3 \ell$ in phenomenological MSSM (pMSSM)}
\subsection{pMSSM with L-violation}

Next we discuss the case of phenomenological MSSM, which includes many
more possibilities than mSUGRA, as far as the mass spectra are
concerned. In particular, the three gaugino mass parameters at the
weak scale then need not be in the approximate ratio $M_1:M_2:M_3 = 1:
2: 6$. Thus the lighter chargino may not be about twice as massive as
the lightest neutralino, a fact that can affect electroweak
phenomenology considerably. Since one of the leptons in the SS$3 \ell$
signal comes from the cascade decay of the chargino (via on or
off-shell $W$'s and sleptons) $\cha \rightarrow \neu l^\pm \nu$, we
need to look into other hierarchies between $M_1$ and $M_2$.

If $M_1 \simeq M_2$, $\cha$, $\tilde \chi_2^0$ and $\neu$ are all very
close in mass, and therefore the lepton coming from chargino decay is
rather soft in the $\cha$ rest frame. But, if the $\cha$ is resulting
from the decay of the gluino or squarks, which could be much heavier,
it can have a large boost, giving rise to high $p_T$ leptons which
will pass the required cuts.

Another interesting situation arises if $M_1 > M_2$. Here, the $\neu$
and the $\cha$ are mostly composed of wino components.  This is what
happens, for example, in the case of anomaly mediated SUSY breaking.
The degeneracy in their masses is even more severe in this case, and
some fine-tuning is necessary to make their mass difference of the
order of pion mass.  Here, one can have an additional channel in the
cascade, from which a third lepton can arise. The second lightest
neutralino can decay to a charged lepton, a neutrino and the $\cha$.
The lepton produced in this way can have sufficient $p_T$ to be
detectable.  The $\cha$, on the other hand, goes to the $\neu$ and an
extremely soft pion (or lepton + neutrino), and the $\neu$ pair can be
the source of same-sign dileptons, which, when of the same-sign as
that of the initial lepton, leads to SS$3 \ell$. We also look into the
case of $M_1<M_2$ in pMSSM but with a wider mass separation than
expected in mSUGRA, namely, $M_2=3M_1$. Here the rates of SS$3 \ell$
are somewhat enhanced. Finally, we look into a kind of
non-universality between the low-energy selectron (or smuon) and stau
soft masses. This can lead to a scenario where all the sleptons except
stau are lighter than $\cha$ and the BF of $\cha$ to leptons is around
$95\%$. Needless to say, this enhances the SS$3 \ell$ rates.

In order to be conservative, we fix the squark soft masses and $M_3$
at $1 \tev$.  We have already shown in Ref.~\cite{SMBM-1} that the
strongly interacting sparticle mass scale of around $600 \gev$ is
easily accessible at the LHC in the SS$3 \ell$ channel during the $7
\tev$ run.  The benchmark points chosen here are just to emphasize
that the SS$3 \ell$ signal can probe a generic MSSM model upto
considerable higher masses of strongly interacting superparticles even
during the early run. In addition, the situation of relatively closely
spaced low-lying charginos/neutralinos, including those with an
inverted hierarchy compared to mSUGRA, also turn up with substantial
event rates. We present the most important parameters in the pMSSM
benchmark points in Table~\ref{tab:pMSSM-BP} and the SS$3 \ell$
cross-sections at these points in Table~\ref{tab:pMSSM-CS}. As
mentioned above, we have fixed the squark soft masses and $M_3$ at $1
\tev$ while $A_t$, $A_b$, $A_{\tau}$ and the $\mu$ parameter have been
fixed at $-500$, $-500$, $-250$ and $975 \gev$ respectively. $\tan \beta$
has been fixed at $10$, and the R-parity violating coupling used for
illustration is $\lambda_{123}=10^{-3}$. The
cross-sections have been calculated for both the $7 \tev$ and the $14
\tev$ runs at the LHC, and we also give the required luminosities for
a five-event discovery, with no events expected from the
backgrounds. The overall usefulness of SS$3 \ell$ in probing low missing-energy
SUSY scenarios is thus brought out quite emphatically by the results
presented by us. To this is added the rather striking prospect of
extracting dynamic information (like the presence of Majorana gauginos
and the exact nature of L-violating couplings) from the SS$3 \ell$ and
SS$4 \ell$ channels, as will be explained in detail in
section~\ref{xy-var}, again on the basis of a model-independent scan
of the SUSY parameter space .

\begin{table}[htb]
 \begin{center}
  \begin{tabular}{|c |c| c| c| c| c| c|}
    \hline
    BP & $M_1$ & $M_2$ & $M_{\neu}$ & $M_{\cha}$ & $M_{\lslep}$ & $M_{\stau}$ \\
       & (GeV) & (GeV) & (GeV)& (GeV)& (GeV)& (GeV)\\
    \hline
    1  & 150 & 150 & 146.54 & 154.80  & 254.13  & 180.91 \\
    2  & 160 & 150 & 154.08 & 154.80  & 254.13  & 217.69 \\
    3  & 100 & 300 & 97.69  & 395.30  & 254.10  & 180.68 \\
    4  & 125 & 250 & 121.65 & 254.35  & 156.76  & 217.52 \\
   \hline
  \end{tabular}

 \end{center}
\caption{Values of $M_1$, $M_2$ and some other relevant parameters 
for the SS$3 \ell$ channel in the pMSSM benchmark points. The squark and
gluino masses are fixed at $\sim 1 \tev$.}
\label{tab:pMSSM-BP}
\end{table}

\begin{table}[htb]
\begin{center}
 \begin{tabular}{|c| c| c| c| c|}
\hline
  BP & $\sigma_{7 \tev}$ & $\mathcal{L}_{7 \tev}$& $\sigma_{14 \tev}$ & $\mathcal{L}_{14 \tev}$\\
     & ($\fb$) & ($\fb^{-1}$)& ($\fb$) & ($\fb^{-1}$)\\
  \hline
   1 &0.91&5.49&4.60&1.09\\
   2 &0.41&12.20&1.62&3.09\\
   3 &2.81&1.78 &20.67&0.24\\
   4 &8.78&0.57&42.93&0.12\\
  \hline
 \end{tabular}

\end{center}
\caption{SS$3 \ell$ cross-sections in the different pMSSM benchmark
  points at $7 \tev$ and $14 \tev$ LHC. We also show the luminosities
  required to obtain $5$ signal events at the two centre of mass
  energies.}
\label{tab:pMSSM-CS}
\end{table}

\subsection{pMSSM with conserved R-parity}

If lepton number is conserved in the MSSM, then it is extremely
difficult to find a scenario where one can obtain a same-sign
trilepton signal (We specifically design the cuts to suppress leptons
coming from b-decays, since otherwise they can boost the standard
model backgrounds as well.)  In fact, we do not find any such scenario
in the simple mSUGRA picture. If one considers a purely
phenomenological MSSM , one can of course generate a wide variety of
mass spectra. We find one particular such spectrum where one can
obtain SS$3 \ell$, but at a negligibly low rate because of branching
fraction suppressions which are difficult to avoid. Thus, as far as we
could analyze the MSSM processes with conserved R-parity, it is not
possible to generate SS$3 \ell$ with significant
cross-section. Therefore, it seems, within a supersymmetric framework,
a reasonably large cross-section of SS$3 \ell$ is a clear indication
of L-violation.

To convince the reader of this, let us outline a scenario in MSSM,
where, in principle, it is possible to obtain an SS$3 \ell$ signal,
albeit with a small rate. Consider a situation where the sbottom is
lighter than the stop. In this case, let us look at stop pair
production ($\stop \stop^*$), followed by the decay $\stop \rightarrow
\sbo W^+$ (and a charge-conjugate decay process for $\stop^*$). The
produced $\sbo$ can then decay to $t {\tilde {\chi_1}}^{-}$, although
with a very low branching fraction. The top quark, of course, then
decays to $b W^+$. Thus starting from the initial $\stop \stop^*$ we
can obtain a final state $(W^+ {\tilde {\chi_1}}^{-} b W^+) (W^-
{\tilde {\chi_1}}^{+} \bar{b} W^-)$. We can re-write this final state
as $(b \bar{b} )(W^+ W^+ {\tilde {\chi_1}}^{+}) (W^- W^- {\tilde
  {\chi_1}}^{-})$. Now, it can be clearly seen that if a set of three
same-charge $W^\pm$'s and $\cha$'s decay leptonically and the other
set decays hadronically, we have a same-sign trilepton signal. In
order to demonstrate the branching fraction suppression of this SS$3
\ell$ final state, let us consider a typical pMSSM spectrum with
$M_{\stop}=522 \gev$ and $M_{\sbo}=482 \gev$. We keep the first two
generation squark masses at $\sim 5 \tev$ in order to separate out the
third generation squark production, which is the only relevant process
for the SS$3 \ell$ channel. The gluino mass is $\sim 1050 \gev$, while
the chargino and neutralino masses are at $264 \gev$ and $120 \gev$
respectively. With these parameters the SS$3 \ell$ cross-section after
all the cuts turns out to be $2.72 \times 10^{-2} \fb$ at the $14
\tev$ LHC, which is evidently very small.

\section{Observable patterns in L-violating LSP decays}
\label{xy-var}
We have now reasons to feel reasonably confident that substantial SS$3
\ell$ (or SS$4 \ell$) rates are unlikely to be seen in R-parity
conserving SUSY, and that R-parity (read lepton number) violation will
be strongly suggested by them. More pointedly, L-violation by odd
units and the existence of more than one Majorana fermions in the
scenario work together towards the enhancement of such signals.

The total rate of SS$3 \ell$ in a particular L-violating scenario
depends not only on the L-violating coupling and the LSP involved, it
is also dictated by SUSY production cross-section and other parameters
determining the cascade decay patterns. We shall now show that {\em it
  is possible to extract the information on the different L-violating
  couplings, through which a Majorana neutralino LSP decays, once we
  make use of the SS$3 \ell$ and SS$4 \ell$ final states.}

With this in view, we construct certain variables which involve not
only the SS$3 \ell$ and SS$4 \ell$ rates in a given scenario, but also
on the total rates in the $3 \ell$ and $4 \ell$ channels. In a generic
MSSM scenario with a particular L-violating coupling, it is possible
to make definite predictions involving these variables based on simple
probability arguments and neutralino branching fraction information in
different combinations of charged lepton final states. We then verify
these predictions using Monte Carlo simulations, where we also show
the effect of selection and isolation cuts, as well as the effect of
adding the SM backgrounds in the $3 \ell$ and $4 \ell$
channels. Although we have demonstrated the results using some mSUGRA
benchmark points for simplicity, the conclusions are generic to
phenomenological scenarios.

\subsection{Neutralino LSP with $\lambda$-type couplings}

A neutralino LSP in presence of $\lambda_{ijk}$-type couplings
contributes to same-sign trileptons only if one of the indices in
$\{ijk\}$ is 3. As $\lambda_{ijk}$ is anti-symmetric in $i$ and $j$,
there are nine independent couplings of $\lambda$-type. Out of these
nine couplings, seven have $3$ as one index, and only two do not have
the index $3$ anywhere. Now consider the generic decay mode of the
$\neu$ where $\neu \rightarrow \tau^{\pm} l^{\mp} \nu \left(l=e,\mu
\right)$ . The produced $\tau^{\pm}$ will decay leptonically in $\sim
35 \%$ of the time, and hadronically in rest of the cases. Now
consider the ratio of the number of same-sign trilepton (SS$3 \ell$)
events to the total number of trilepton events (which includes both
same-sign trileptons (SS$3 \ell$) and mixed-sign trileptons (MS$3
\ell$)). This ratio can be calculated independent of the other SUSY
parameters as follows. In the above case, in a trilepton event, we
know that at most one of the leptons is coming from the cascade as the
pair of neutralinos produced at the end of the decay chains will
always give rise to at least 2 leptons. As mentioned before, the
produced $\tau^\pm$'s decay to a semi-leptonic final state in $\sim
35\%$ of the cases, and to hadronic final states in $65 \%$
cases. Therefore, as the two $\neu$ decays will produce two leptons
when both the $\tau^\pm$'s decay hadronically, the fraction of cases a
pair of $\neu$'s goes to 2 leptons and jets and neutrinos is $(0.65)^2
= 0.4225$. Similarly, a pair of neutralinos can go to 3 leptons in a
fraction $\left (2 \times 0.65 \times 0.35 \right)=0.455 $ of all
cases (i.e., when one $\tau^\pm$ decays leptonically and the other one
decays hadronically). In rest of the cases they decay to a four-lepton
final state (when both the $\tau^\pm$'s decay leptonically) which we
are not considering in this case. Thus out of all possible trilepton
events, in $\simeq 42 \%$ cases one lepton comes from the cascade and
in $\sim 46 \%$ cases no lepton comes from the cascade. We can
summarise the situation in Table~\ref{tab1}.

\begin{table}[ht]
	\centering

\begin{tabular}{|c| c| c| c| c|}

\hline
No. of leptons &No. of leptons & Fraction & SS$3 \ell$ Fraction & MS$3 \ell$ Fraction\\
(Cascade)      &(LSP Decay)    & of cases &               &              \\
\hline

1  & 2 & 0.42  & 0.25 & 0.75 \\

0 & 3 & 0.46  & 0 & 1 \\

\hline

	\end{tabular}

\caption{\label{tab1} Fraction of trilepton events with different
  origins for the leptons, and the fractions of SS$3 \ell$ and MS$3
  \ell$ events among them (see explanation in text).}
\end{table}

In Table~\ref{tab1}, the first two columns represent the number of
leptons coming from the two different sources that we distinguish,
namely, from the cascade and from the decay of the two $\neu$
LSP's. As explained above, there are only two such possibilities in a
trilepton event. Those two possibilities are described in the two rows
of the table. The third column describes the fraction of cases in
which each of these possibilities occur. We have explained the numbers
in this column above. Finally, the last two columns represent the
fraction of SS$3 \ell$ and MS$3 \ell$ events in each of the possible
ways of obtaining a trilepton event, as explained below.

From Table~\ref{tab1}, we see that in the first case where two of the
leptons come from the LSP-pair decay, and one from the cascade, the
probability of getting an $l^+l^+$ pair from the LSP's is $0.25$, and
same for obtaining an $l^-l^-$ pair (this stems from the fact that the
$\neu$ is Majorana). Now, in a trilepton event, let $P_1$ be the
probability of the single lepton coming from the cascades being of
positive charge, and $P_2$ for it to be of negative charge. Then, the
probability of obtaining an SS$3 \ell$ event is $0.25 \times P_1 +
0.25 \times P_2 = 0.25$ as $P_1+P_2 =1$. Therefore, the probability of
obtaining an MS$3 \ell$ event is $1-0.25 =0.75$. In the second case,
where all three of the leptons are coming from the LSP decay, all the
trilepton events are of MS$3 \ell$-type. Note that, we are demanding
only three leptons in the final state, therefore any event with
additional leptons (four or more) are vetoed out. Let us now define
the ratio

\begin{equation}
\label{eqn:x}
 x = \frac{\sigma_{SS3 \ell}}{\sigma_{SS3 \ell}+\sigma_{MS3 \ell}}
\end{equation}

From Table~\ref{tab1}, we see that we can easily calculate this ratio as follows
\begin{equation}
 x  = \frac{\sigma_{total}\times 0.42
 \times 0.25}{\sigma_{total}\times \left [\left (0.42 \times 0.25 \right)+\left(0.42 \times 0.75 +
 0.46 \right)\right]} \simeq 0.12 ,
\end{equation}

where $\sigma_{total}$ is the total SUSY production cross-section
which cancels out among the numerator and the denominator. As we have
explained above, the trilepton events are a fraction of all possible
SUSY events and SS$3 \ell$ and MS$3 \ell$ events are subsets of all $3
\ell$ events.  This value of the ratio $x \simeq 0.12$ is therefore a
prediction stemming from the L-violating decay mode of the neutralino
under study and also the Majorana character of the
neutralino. Whatever be the values of the other SUSY parameters, as
long as we have a $\neu$ LSP decaying via a $\lambda_{123}$ type of
coupling, this ratio is fixed. In particular, this ratio is
independent of the probabilities of obtaining a charged lepton of
either sign from the cascade. One should note, however, that if the
L-violating couplings are so large as to compete with the gauge
couplings for the decay of sparticles other than the LSP, this result
can change. But, flavour physics and neutrino physics experiments
suggest that these Yukawa couplings would take rather small values if
SUSY models are to explain the above phenomena.

\begin{table}[htb]
\begin{center}
\begin{tabular}{|l|c|c|c|}
\hline
Point & $\sigma_{SS3 \ell}+\sigma_{MS3 \ell}$ & $\sigma_{SS3 \ell}$ & $x$  \\
      & (fb)                          &(fb)             &      \\
\hline
1a(1)&928.32&75.11&0.08\\
1a(2)&1084.26&110.06&0.10\\
1a(3)&228.24&27.51&0.12\\
1a(4)&149.47&14.20&0.10\\
\hline
\end{tabular}
\end{center}
\caption{\label{x-values} The $3 \ell$ (signal+SM background) and SS$3
  \ell$ (signal) cross-sections at $14 \tev$ LHC after the
  $m_{eff}>250\,\gev$ cut and the ratio $x$ calculated including the
  SM background contribution. The total SM background in the $3 \ell$
  channel after the above cut is $10.12 \fb$. Note that the predicted
  value of $x$ in this case is $\sim 0.12$, which shifts somewhat
  after including the effects of lepton isolation, detection
  efficiencies, other cuts and the SM backgrounds. The agreement with
  the predicted value is within $20 \%$ in most cases.}
\end{table}

In realistic situations, where we have to consider the experimental
triggers, detector efficiencies etc., the ratio $x$ can fluctuate
around the predicted value of $\sim 0.12$. Unfolding these effects in
an event by event basis is not an easy exercise, and we abstain from
trying to do so. In Table~\ref{x-values} we present the ratio $x$
obtained by Monte Carlo simulations with proper cuts in different
benchmark points with widely varying SUSY parameters. We see that to
within $20 \%$ one always gets a ratio as predicted, thereby
validating the above analysis. Thus we find that this ratio of SS$3
\ell$ to the total trilepton production cross-section gives us dynamic
information about the underlying SUSY theory, in particular the
L-violating coupling involved and the Majorana nature of the decaying
LSP.

What happens if we change the L-violating coupling? Note that, in the
presence of a generic $\lambda_{ijk}$-type coupling a $\neu$ decays to
two charged leptons and a neutrino. Only if one of these leptons is a
tau, which in turn can decay hadronically, one can obtain an SS$3
\ell$ signal. Therefore, one of the indices in $\{ijk\}$ has to be
$3$. Moreover, we find the ratio $x \simeq 0.12$ only for the
none-zero coupling $\lambda_{123}$. The reason for this is that if
$i=3$ or $j=3$ (which are equivalent due to the antisymmetry of
$\lambda_{ijk}$ in the indices $i,j$), then the $\neu$ can also decay
to a $\nu_{\tau}$ instead of a $\tau^{\pm}$, thereby changing the
ratio. For example, for the set of couplings $\{131, 132, 231, 232 \}$
we find this ratio to be approximately $x \sim 0.14$, while for the
set $\{121, 122 \}$, $x=0$, since no SS$3 \ell$ events are expected in
these cases .

The above analysis thus shows that the dynamic information of a
Majorana $\neu$ decaying via $L_iL_jE^c_k$-type couplings can be
captured in a quantity easily measurable at the LHC experiments. Since
the cross-sections in the $3 \ell$ and SS$3 \ell$ channels are rather
large (see Table~\ref{x-values}) at the $14 \tev$ LHC, one can acquire
a reasonably good statistics within $1-5 \fb^{-1}$ of integrated
luminosity. Therefore, these cross-sections can be measured and the
ratio $x$ calculated fairly accurately in the early periods of the $14
\tev$ run.

\subsection{Neutralino LSP with $\lambda^\prime$-type couplings} 

The case for $\neu$ LSP with $\lambda^\prime$-type couplings is
somewhat more complicated than that for the $\lambda$-type couplings.
No unique prediction (which is independent of the other SUSY
parameters) can be made there about the ratio $x$. One can, however,
construct a similar ratio with four-lepton events. Subsequently, we
obtain a linear relation between these two ratios, which is then
independent of the parameters determining the cascade decays.

In the presence of a $\lambda^\prime$-type coupling, a $\neu$ decays
either to two quarks and a neutrino, or to two quarks and a charged
lepton. SS$3 \ell$ signals can arise in the second case. But now we
have more ways in which one can obtain trilepton events. Let us define
the fraction of cases in which a $\neu$ decays via $\neu \rightarrow
l^\pm q \prime q$ to be $\alpha$. Now let us note the various possible
ways of obtaining trilepton events in Table~\ref{neu-lamdap}. The
structure and meaning of the different entries in this table are same
as explained in detail for Table~\ref{tab1}.

 \begin{table}[ht]
	\centering

\begin{tabular}{|c| c| c| c| c|}

\hline
No. of leptons &No. of leptons & Fraction & SS$3 \ell$ Fraction & MS$3 \ell$ Fraction\\
(Cascade)      &(LSP Decay)    & of cases &               &              \\
\hline
3  & 0 & $\left(1-\alpha \right)^2$  &0  & 1 \\

2  & 1 &$2 \alpha \left(1-\alpha \right)$   &$\frac{1-P_3}{2}$  &$\frac{1+P_3}{2}$  \\

1  & 2 & $\alpha^2$  & 0.25 & 0.75 \\

\hline

	\end{tabular}

\caption{\label{neu-lamdap} Fraction of trilepton events with
  different origins for the leptons, and the fractions of SS$3 \ell$
  and MS$3 \ell$ events among them (see explanation in text).}
\end{table}

Since $\alpha$ denotes the probability that a $\neu$ will decay
leptonically, the fraction of cases a pair of $\neu$'s give rise to
two leptons is $\alpha^2$. Similarly, when both the $\neu$'s decay to
neutrinos and quarks, we do not obtain any leptons from LSP
decays. This happens in $\left(1-\alpha \right)^2$ fraction of
trilepton events. And, finally, in the remaining $2 \alpha
\left(1-\alpha \right)$ fraction of cases, we obtain one lepton from
the decay of the two LSP's. In the first case, when all three of the
leptons come from the cascade, we do not obtain any SS$3 \ell$ event,
making the MS$3 \ell$ fraction unity. In order to understand the
second case, note that in Table~\ref{neu-lamdap}, when two leptons
come from the cascade, we define $P_3$ to be the probability of them
being oppositely charged ($l^\pm l^\mp$). Thus, $\left ( 1-P_3
\right)$ is the probability of them being of same charge ($l^\pm
l^\pm$). In such a case, in a trilepton event, evidently the third
lepton comes from LSP decay.  In half of such events the lepton coming
from LSP decay will also have the same sign, thereby giving rise to an
SS$3 \ell$ event. Thus the probability of obtaining an SS$3 \ell$
event is $\frac{1-P_3}{2}$. Consequently, the probability for
obtaining a MS$3 \ell$ event is $\left (1- \frac{1-P_3}{2} \right) =
\frac{1+P_3}{2}$. In the third case, where two of the leptons come
from LSP decay, because of the Majorana nature of the decaying $\neu$
LSP, we get the corresponding fractions in the same way as in the
previous sub-section, where we considered $\neu$ decay via
$\lambda$-type terms.

In this case, therefore, we find the following formula for the ratio
$x$ defined in eqn.~\ref{eqn:x}.

\begin{equation}
 x =  \alpha - \frac{3}{4} \alpha^2 -P_3 \left(\alpha-\alpha^2 \right)
\end{equation}

Now, the ratio $\alpha$ is very weakly dependent on the sparticle mass
spectra, especially the difference between the up and down-type squark
masses entering the off-shell propagators in the 3-body $\neu$
decays. In most scenarios this difference is rather small, especially
for the first two families.  On the whole, $\alpha$ is close to 0.5 in
most cases.

As mentioned before, there is a residual dependence of $x$ on $P_3$,
thereby making this ratio vary as the other SUSY parameters vary (also
the parton distribution functions affect $P_3$). In order to eliminate
$P_3$ and obtain a prediction that follows just from the Majorana
nature of the $\neu$ and the L-violating coupling involved, we
introduce another ratio $y$ defined as

\begin{equation}
\label{eqn:y}
 y = \frac{\sigma_{SS4 \ell}}{\sigma_{SS4 \ell}+\sigma_{MS4 \ell}}
\end{equation}

where $\sigma_{SS4 \ell}$ and $\sigma_{MS4 \ell}$ are the same-sign
four-lepton and mixed-sign four-lepton cross-sections respectively.

To calculate $y$, we make a table similar to the one made for calculating $x$.

\begin{table}[ht]
	\centering

\begin{tabular}{|c| c| c| c| c|}

\hline
No. of leptons &No. of leptons & Fraction & SS$4 \ell$ Fraction & MS$4 \ell$ Fraction\\
(Cascade)      &(LSP Decay)    & of cases &               &              \\
\hline
%
%
4  & 0 & $\left(1-\alpha \right)^2$  &0  & 1 \\

3  & 1 &$2 \alpha \left(1-\alpha \right)$   & 0 &1  \\

2  & 2 & $\alpha^2$  & $\frac{1-P_3}{4}$ &$\frac{3+P_3}{4}$\\

\hline
	\end{tabular}

\caption{\label{neu-lamdap-4l} Fraction of four-lepton events with
  different origins for the leptons, and the fractions of SS$4 \ell$
  and MS$4 \ell$ events among them (see explanation in text).}
\end{table}

Since, the fraction of cases for the different possibilities are only
dependent on the number of leptons coming from LSP decays, the entries
in the third column of Table~\ref{neu-lamdap-4l} can be understood in
the same way as in the trilepton case, which we explained before while
discussing Table~\ref{neu-lamdap}. In the first two cases, where four
and three leptons come from the cascade respectively, the MS$4 \ell$
fraction is 1, since we cannot get more than two same-sign leptons
from the cascade. In the third case, we define $P_3$ as before. Since
$\left(1-P_3\right)$ is the probability to have a same-sign lepton
pair from the cascade, in order to obtain a same-sign four lepton
event, we need the other two leptons coming from LSP decay to be of
the same-sign as that of the cascade leptons. Now, since the $\neu$ is
a Majorana particle, when it decays leptonically, the probability to
obtain a same charge lepton as in the cascade is $1/2$, and similarly
for the second $\neu$, thus giving us a probability of
$\frac{1-P_3}{4}$ to obtain an SS$4 \ell$ event. The rest of the
events are of MS$4 \ell$ variety, which come with a fraction of $\left
(1-\frac{1-P_3}{4}\right)=\frac{3+P_3}{4} $. This completes the
explanation of Table~\ref{neu-lamdap-4l}.

From Table~\ref{neu-lamdap-4l} and eqn.~\ref{eqn:y} we find that

\begin{equation}
 y= \frac{\alpha^2}{4} - \frac{\alpha^2 P_3}{4}.
\end{equation}

The total SUSY production cross-section $\sigma_{total}$ cancels out
in the ratio as in the case for $x$.  Combining these two equations
for $x$ and $y$, we can eliminate $P_3$ to obtain the following
equation relating $x$ and $y$:

\begin{equation}
\label{x-y}
 x = \frac{\alpha^2}{4}+4y \left (\frac{1}{\alpha}-1\right)
\end{equation}

This equation is therefore a prediction based just on the Majorana
nature of the decaying $\neu$ LSP and the presence of
$\lambda^\prime$-type couplings. In order to verify the above claim
and also to see the deviations due to lepton selection and isolation
effects we note the values of $x$ and $y$ obtained in different
benchmark points and compare them with the above prediction taking
$\alpha \sim 0.5$ and present the results in Table 11.

\begin{table}[htb]
\begin{center}
\begin{tabular}{|l|c|c|c|c|c|c|}
\hline
Point & $y^{S}_{MC}$ & $x^{S}_{MC}$ & $x^{S}_{eqn.}$ & $y^{S+B}_{MC}$ & 
$x^{S+B}_{MC}$ & $x^{S+B}_{eqn.}$  \\
\hline
2(1)&0.024 &0.147 &0.159 &0.018 &0.111 &0.135\\
2(2)&0.023 &0.154 &0.155 &0.020 &0.126 &0.143\\
2(3)&0.027 &0.144 &0.171 &0.021 &0.102 &0.147\\
2(4)&0.044 &0.175 &0.239 &0.025 &0.094 &0.163\\
\hline 
\end{tabular}
\end{center}
\caption{\label{x-y-lp} The ratios $x$ and $y$ after the
  $m_{eff}>250\,\gev$ cut, before and after adding the SM background
  cross-sections. Here, $y^{S}_{MC}$ and $x^{S}_{MC}$ refer to the
  ratios $x$ and $y$ calculated only with the signal whereas
  $y^{S+B}_{MC}$ and $x^{S+B}_{MC}$ denote the ratios calculated
  adding up both the signal and the background cross-sections in the
  appropriate channels.  $x^{S}_{eqn.}$ and $x^{S+B}_{eqn.}$ denote
  the values of $x$ calculated using eqn.~\ref{x-y} taking
  $\alpha=0.5$, with $y^{S}_{MC}$ and $y^{S+B}_{MC}$ as the respective
  inputs.}
\end{table}

The entries of Table~\ref{x-y-lp} have been explained in the caption
of the table. The ratios $x$ and $y$ have been evaluated from
cross-sections calculated for the $14 \tev$ LHC.  As the total $3
\ell$ and $4 \ell$ cross-sections in the case of a $\neu$ LSP with
$\lambda^{\prime}$-type coupling are comparable to the SM backgrounds,
the ratios $x$ and $y$ change after adding the backgrounds. In order
to show that we present the ratios both before and after adding the SM
background cross-sections. In order to validate the prediction derived
in eqn.~\ref{x-y}, we take the Monte Carlo prediction for $y$ as an
input, and then calculate the value of $x$ from eqn.~\ref{x-y}, and
denote it by $x_{eqn.}$. This $x_{eqn.}$ is then compared with
$x_{MC}$, the value obtained from Monte Carlo. The rather excellent
agreement between the entries in the third and fourth columns in
Table~\ref{x-y-lp} demonstrates the viability of our claim in
eqn.~\ref{x-y}. As noted before, here we have used the approximate
value of $0.5$ for $\alpha$. We then again repeat the same calculation
after adding the SM backgrounds in the $3 \ell$ and $4 \ell$
channels. Since the backgrounds are comparable to the signal in this
case, the ratios change somewhat after the background addition, and
the agreement between the prediction of eqn.~\ref{x-y} and the MC is
not as good as with only the signal, which is expected. Also note
that, in the benchmark points $2(2)$ and $2(4)$ we have the $\cha$
lighter than the $\lslep$. This reduces the $\cha$ branching fraction
to leptons, thereby leading to a reduction of trilepton and
four-lepton events of same-sign and mixed-sign varieties. The lower
branching fraction is compensated by the much larger total SUSY
production cross-section in point $2(2)$. Point $2(4)$ thus suffers
from lower number of multi-lepton events, and the accurate evaluation
of the ratios $x$ and $y$ here would require much larger
statistics. Also if SS$3 \ell$ signals are indeed seen in the mass
range of, say point $2(4)$, then the further reduction of backgrounds
may be a pressing need in order to extract dynamics out of this
signal.
  
\subsection{Neutralino LSP with bi-linear couplings} 
In the presence of bi-linear L-violating couplings, the decay
branching fractions of $\neu$ in different channels are dependent on
various other soft SUSY-breaking parameters, too. Thus it is not
possible to predict a specific equation which will be valid
generically for all possible choices of the relevant
parameters. Instead, we focus in a region where the $\neu$ decays
either in the $W^{\pm} \mu^\mp / \tau^\mp$ or in the $Z \nu$
channel. This is largely the case when the slepton/sneutrino states
have not-too-large mixing with the Higgs states~\cite{R_M_2}.  In this
case, we find an equation relating the $x$ and $y$-variables which is
very similar to the equation of straight line found for the case of
$\neu$ LSP with $\lambda^\prime$-type couplings (with a different
slope for the straight line!). Since the detailed evaluation of the
relevant branching fractions in different combinations of
charged-lepton final states is straightforward but cumbersome, we just
note down the final results in the following tables.

 \begin{table}[ht]
	\centering

\begin{tabular}{|c| c| c| c| c|}
\hline

No. of leptons &No. of leptons & Fraction & SS$3 \ell$ Fraction & MS$3 \ell$ Fraction\\
(Cascade)      &(LSP Decay)    & of cases &               &              \\
\hline

3  & 0 & 0.146  &0  & 1 \\

2  & 1 &0.366  &$\frac{1-P_3}{2}$  &$\frac{1+P_3}{2}$  \\

1  & 2 &0.335  & 0.17 & 0.83 \\

0  & 3 &0.134  &0     &1    \\

\hline
	\end{tabular}

\caption{\label{neu-bilinear} Fraction of trilepton events with
  different origins for the leptons, and the fractions of SS$3 \ell$
  and MS$3 \ell$ events among them, in the case of a $\neu$ LSP with
  bi-linear L-violating couplings (see explanation in text).}
\end{table}
\begin{table}[htb!]
	\centering

\begin{tabular}{|c| c| c| c| c|}

\hline
No. of leptons &No. of leptons & Fraction & SS$4 \ell$ Fraction & MS$4 \ell$ Fraction\\
(Cascade)      &(LSP Decay)    & of cases &               &              \\
\hline

4  & 0 & 0.146  &0  & 1 \\

3  & 1 &0.366  & 0 &1  \\

2  & 2 & 0.335 &$0.17-0.17P_3$&$0.83+0.17P_3$\\

1  & 3 & 0.134 &0 &1 \\

0  & 4 & 0.020 &0 &1 \\

\hline

	\end{tabular}

\caption{\label{neu-bilinear-4l} Fraction of four-lepton events with
  different origins for the leptons, and the fractions of SS$4 \ell$
  and MS$4 \ell$ events among them, in the case of a $\neu$ LSP with
  bi-linear L-violating couplings (see explanation in text).}
\end{table}

From Table~\ref{neu-bilinear} we find that

\begin{equation}
 x= 0.24 - 0.19P_3
\label{bi-x}
\end{equation}

and similarly, from Table~\ref{neu-bilinear-4l} we find an expression for $y$

\begin{equation}
 y = 0.06 (1-P_3)
\label{bi-y}
\end{equation}

We can eliminate $P_3$ from equations~\ref{bi-x} and~\ref{bi-y}, to
obtain an equation of straight line relating $x$ and $y$

\begin{equation}
\label{eqn:x-y-bi}
 x = 3.529y + 0.063
\end{equation}

As mentioned before, this equation is very similar to the equation
obtained for the case of $\neu$ LSP with $\lambda^\prime$-type
couplings. The slope in the $x-y$ plane, however, is slightly
different in this case.

\begin{table}[htb]
\begin{center}
\begin{tabular}{|l|c|c|c|c|c|c|}
\hline
Point & $y^{S}_{MC}$ & $x^{S}_{MC}$ & $x^{S}_{eqn.}$ & $y^{S+B}_{MC}$ & 
$x^{S+B}_{MC}$ & $x^{S+B}_{eqn.}$  \\
\hline
3(1)&0.012 &0.096 &0.107 &0.012 &0.088 &0.105\\
3(2)&0.016 &0.087 &0.112 &0.015 &0.075 &0.115\\
\hline 
\end{tabular}
\end{center}
\caption{\label{x-y-bil}
Same as in Table~\ref{x-y-lp}, for the case of bi-linear L-violation.}
\end{table}

In Table~\ref{x-y-bil}, which is similar to Table~\ref{x-y-lp}, we
present a comparison of the MC calculation and the prediction from
eqn.~\ref{eqn:x-y-bi} for x, with $y$ calculated from MC as input. We
find that the predictions agree with the MC calculations to within
$\sim 20 \%$ or better. As for the $\lambda^\prime$ case, the
prediction and MC calculations deviate a little bit more after adding
the SM backgrounds, since the total signal cross-sections in the $3
\ell$ and $4 \ell$ channels are comparable to the backgrounds. Also,
in this case as the total rates for multi-lepton events are rather
small for the chosen benchmark points (with the squark and gluino
masses $\sim 1 \tev$), we need a much larger statistics in order to
calculate the ratios accurately.

\section{Summary and conclusion}
We have performed a detailed study of SS$3 \ell$ and SS$4 \ell$
signals in the context of the LHC, to arrive at a number of important
conclusions.  First, such signals are enhanced, to such a degree as to
be appreciable even during the 7 TeV run (and also the 14 TeV run with
low integrated luminosity), if there is (a) L-violation by odd units,
and (b) the presence of self-conjugate fields. The outstanding
theoretical scenario meeting the above requirements is SUSY with
R-parity violated via lepton number.  Therefore, we strongly advocate
the investigation of such signals, especially as they are
complementary to signals with large missing $E_T$.

We have gone beyond the mSUGRA scenario and focused on different
regions of the parameter space of a general SUSY model. It has been
shown that sizable SS$3 \ell$ rates are expected over various regions
of interest in the parameter space, so much so that upto a TeV in the
scale of strongly interacting superparticle masses can be explored at
the 7 TeV run itself. This in itself is quite remarkable for signals
with such high multiplicity of leptons, and can be attributed to
almost non-existent SM backgrounds. It is further shown that event
rates of comparable magnitude are almost impossible to achieve in an
L-conserving SUSY scenario of a general kind. This, we argue, further
strengthens the motivation of studying same-sign multileptons.

The other really useful feature of SS$3 \ell$ and SS$4 \ell$ signals
that we have emphasized is that they enable us to extract information
on the dynamics of R-parity violation, namely, whether lepton number
is violated through the $\lambda$, $\lambda^\prime$ or the bilinear
terms. Using SS$3 \ell$ and SS$4 \ell$ event rates in conjunction with
their mixed-sign counterparts, one is able to define certain ratios
and their relationships which are typical of the type of R-parity
violating terms, taken one at a time. More importantly, these ratios
and their relations are largely independent of the SUSY spectrum and
the nature of cascades, and depend centrally on the Majorana character
of neutralinos, making our conclusions extremely general. We perform
detailed simulation for a number of benchmark points to substantiate
this claim. The simulations include the effects of experimental cuts
as well the SM backgrounds for mixed-sign trileptons and four leptons.

Thus the overwhelming recommendation is for a careful analysis of SS$3
\ell$ and SS$4 \ell$ signals at the LHC. Such analysis should be
concurrent with the search for events with large missing energy,
because of its complementary nature.

\section*{Acknowledgments} 
We thank Sanjoy Biswas and Kaoru Hagiwara for useful discussions.
This work was partially supported by funding from the Department of
Atomic Energy, Government of India for the Regional Centre for
Accelerator-based Particle Physics, Harish-Chandra Research Institute
(HRI). We also acknowledge the cluster computing facility at HRI
(http:/$\!$/cluster.hri.res.in).

\end{document}